\documentclass{KapProc} 

\normallatexbib
\let\lcitebracket[
\let\rcitebracket]
\makeatletter \renewcommand\@biblabel[1]{{#1}. } \makeatother
\usepackage{epsf}
\usepackage{here}

\begin{document}
\begin{sloppy}

\articletitle[Nonlinear Impurity Modes in Homogeneous and Periodic Media]{Nonlinear Impurity Modes in\\ Homogeneous and Periodic Media}

\author{Andrey A. Sukhorukov and Yuri S. Kivshar}
\affil{Nonlinear Physics Group, RSPhysSE\\
Australian National University\\
Canberra ACT 0200, Australia}
\email{ans124@rsphysse.anu.edu.au,~ysk124@rsphysse.anu.edu.au}

\begin{abstract}
We analyze the existence and stability of nonlinear localized waves described by the Kronig-Penney model with a nonlinear impurity. We study the properties of such waves in a homogeneous medium, and then analyze new effects introduced by periodicity of the medium parameters. In particular, we demonstrate the existence of a novel type of stable nonlinear band-gap localized states, and also reveal an important physical mechanism of the oscillatory wave instabilities associated with the band-gap wave resonances.
\end{abstract}


\section{Introduction}

Wave propagation in periodic media is associated with many interesting physical phenomena~\cite{yeh,book}. Modern technology allows to create different kinds of macro- and mesoscopic periodic and layered structures such as semiconductor superlattices and heterostructures, magnetic multilayers possessing the giant magnetoresistance, multiple-quantum-well structures, optical waveguide arrays, photonic band-gap materials and photonic crystal fibers, etc. The main feature of different periodic structures (which follows from the classical Floquet-Bloch theory) is the existence of forbidden frequency band gaps (or stop bands) where {\em linear waves undergo Bragg reflection} from the periodic structure~\cite{yeh}. However, many of the recently fabricated periodic structures exhibit pronounced nonlinear properties that give rise to qualitatively new physical effects such as {\em multistability} of a finite nonlinear periodic medium~\cite{multi} and energy localization in the form of {\em gap solitons}~\cite{gap}. Such effects are usually analyzed in the framework of the coupled-mode theory~\cite{coupled}, and they are associated with the nonlinearity-dependent tuning of the stop-band as the wave intensity is increased.

Another fundamental reason for the wave localization is the presence of impurities. As a matter of fact, artificially introduced inhomogeneities are often used as a powerful means of controlling wave scattering. However, the impurity-induced localization in {\em nonlinear and periodic media} is largely an open area of research. We stress that the impurity response is determined by the local field amplitude, and therefore the standard averaging procedure can not be directly applied to this kind of problems.

Recent experiments, e.g. the observation of optical gap solitons~\cite{gap_exp} and the control of coherent matter waves in optical lattices~\cite{kasevich}, as well as theoretical results such as the discovery of the oscillatory instability of gap solitons~\cite{gap_inst}, call for a systematic analysis of nonlinear effects in periodic structures and band-gap localized states {\em beyond the approximation provided by the coupled-mode theory}. As we show below, such an analysis is crucially important for determining {\em stability of nonlinear waves in periodic structures} because the wave instabilities can appear due to the mode coupling to other bands.

In this paper, we analyze {\em the existence and stability of nonlinear localized waves in a periodic medium with multiple gaps in the transmission spectrum} that is valid beyond the coupled-mode theory. We consider a simple model where waves are localized in a layered medium by an intensity-dependent impurity (or, in other words, they are guided by a thin-layer nonlinear waveguide). Assuming the applicability of our results to a variety of different physical systems (see below), we  characterize, in the framework of a unified and systematic approach, the properties of {\em two types of nonlinear localized waves}~\cite{ref}. In particular, we reveal an important physical mechanism of wave instability associated with the band-gap resonances. We demonstrate also that several types of {\em stable band-gap localized states} can exist in the presence of nonlinearity.

\section{General analysis}
\subsection{Model equations}

We consider a general problem in which the dynamics of elementary excitations of a physical system (e.g., phonon, magnons, etc.) is described by an effective equation for the wave-packet envelope $\psi (x,t)$~\cite{kosevich}. When the density of such quasi-particles becomes high enough, their interaction should be taken into account, e.g. in the framework of the mean-field approximation. In the simplest case, the quasi-particle interaction and collective phenomena in an inhomogeneous medium can be described by the nonlinear Schr\"odinger (NLS) equation for the wave-packet envelope $\psi (x,t)$,
\begin{equation} \label{eq:nls}
     i \frac{\partial \psi}{\partial t} 
     + \frac{\partial^2 \psi}{\partial x^2}  
     + {\cal F}(I; x) \psi = 0,
\end{equation}
where $I \equiv |\psi|^2$ is the wave intensity, $t$ is time (or propagation variable), $x$ is the spatial coordinate, and the real function ${\cal F}(I; x)$ describes both {\em nonlinear} and {\em periodic} properties of the medium. We note that the system~(\ref{eq:nls}) is Hamiltonian, and for localized solutions the power $P = \int_{-\infty}^{+\infty} I(x)\; dx$ is finite, and it is a conserved quantity.

We assume that the superlattice is linear, and nonlinearity appears only through the intensity dependent response of an embedded localized impurity. Then, if the corresponding width of the wave envelope is much larger than that of the impurity, the inhomogeneity can be modeled by a delta-function and, in the simplest case, we can write 
\begin{equation} \label{eq:def_F}
   {\cal F}(I; x) = \nu(x) + \delta(x) G(I), 
\end{equation}
where the function $G(I)$ characterizes the properties of the impurity, and $\nu(x) \equiv \nu(x+h)$ describes an effective potential of the superlattice with the spatial period $h$. 

Model~(\ref{eq:nls}),(\ref{eq:def_F}) appears in different physical problems of the macroscopic nonlinear dynamics of solids and nonlinear optical systems. In particular, it describes a special case of a more general problem of the existence and stability of nonlinear guided waves in a layered (or stratified) dielectric medium, where the delta-function impurity corresponds to a very thin layer embedded into a nonlinear medium with a Kerr or non-Kerr response~\cite{optics}; in this case the time variable $t$ stands for the propagation coordinate along the layer, and $x$ is the transverse coordinate. Model~(\ref{eq:nls}),(\ref{eq:def_F}) can also appear in the problem of the Bose-Einstein condensation in optical lattices~\cite{kasevich}.

\subsection{Localized modes}

We study stationary localized solutions of Eq.~(\ref{eq:nls}) in the form $\psi(x,t) = u(x) e^{i \omega t}$, where $\omega$ is the normalized frequency (or the propagation constant, in optics), and the real function $u(x)$ satisfies the equation:
\begin{equation} \label{eq:u0_inh}
  - \omega u + \frac{d^2 u}{d x^2} + {\cal F}(I; x) u = 0.
\end{equation}
For a local nonlinearity described by Eq.~(\ref{eq:def_F}), the localized waves can be constructed with the help of a matching condition, using the solutions of Eq.~(\ref{eq:u0_inh}) with ${\cal F}(I; x) = \nu(x)$, presented in the form of the Bloch-type wave functions~\cite{yeh}. 

If the effective periodic potential $\nu(x)$ is approximated by a piecewise-constant function (as in the case of the so-called Kronig-Penney model), the solution on both sides of the impurity can be decomposed into a pair of counter-propagating waves with the amplitudes $a(x)$ and $b(x)$,
\[
  u_b(x) = a(x) e^{- {\mu}(x) x} 
         + b(x) e^{{\mu}(x) x}, 
\]
where $\mu(x) = \sqrt{\omega - \nu(x)}$ is the local wavenumber.
As follows from the Floquet-Bloch theory, for a Bloch-wave solution the reflection coefficient $r(x) = b(x) / a(x)$ is a periodic function, i.e. $r(x) = r(x + h)$, and it is found from the following eigenvalue problem:
\begin{equation} \label{eq:eigen}
   T(x) \left( \begin{array}{l} 1 \\ r(x) \end{array} \right) 
    =
      \tau(x) 
      \left( \begin{array}{l} 1 \\ r(x) \end{array} \right) ,
\end{equation}
where $T(x)$ is a {\em transfer matrix} that describes a change of the wave amplitudes $\{a,b\}$ after one period $(x,x+h)$. An explicit expression for the transfer matrix can be found in Ref.~\cite{our_pre}, where it was also proven that ${\rm det}\;T \equiv 1$, and two linearly independent solutions of Eq.~(\ref{eq:eigen}) correspond to a pair of the eigenvalues $\tau$ and $\tau^{-1}$. Relation $\tau(\omega)$ determines a {\em band-gap structure} of the superlattice spectrum: the waves are {\em propagating}, if $|\tau| = 1$, and they are {\em localized}, if $|\tau| \ne 1$. In the latter case, a nonlinear impurity can support {\em nonlinear localized waves}, and 
the wave amplitude at the impurity is determined from the continuity condition~\cite{our_pre} at $x=0$ [$I_0 \equiv I(0)$]:
\begin{equation} \label{eq:zeta}
  G_0 \equiv G(I_0) = \zeta(\omega) .
\end{equation}
Here $\zeta = (\zeta^+ + \zeta^-)_{x=0}$, 
${\zeta^\pm} = \mu^\pm {\left( 1 - r^\pm \right)} 
                       {\left( 1 + r^\pm \right)}^{-1}$, 
and we denote with ``$+$'' and ``$-$'' the lattice characteristics on the right and the left side of the nonlinear impurity, respectively, i.e. $\nu(x) = \nu^+(|x|)$, for $x>0$, and $\nu(x) = \nu^-(|x|)$, for $x<0$. Note that the reflection coefficients should correspond to the Bloch-wave solutions with the asymptotics $|u( x \rightarrow \pm \infty )| \rightarrow 0$. 

Relation~(\ref{eq:zeta}) allows us to identify different types of nonlinear localized states. We notice that such localized states, being supported by an attractive nonlinear impurity ($G_0 > 0$), exist in the so-called {\em waveguiding} regime, when $\zeta(\omega) > 0$. Additionally, localization can occur at a repulsive nonlinear impurity ($G_0 < 0$) in the {\em anti-waveguiding} regime, provided $\zeta(\omega) < 0$.

\subsection{Stability properties}

To study {\em the linear stability}, we consider the evolution of small-amplitude perturbations of the localized state presenting the solution in the form
\[
 \psi (x,t) 
 = \left\{ u(x) + v(x) e^{i \gamma t} 
                + w^{\ast}(x) e^{-i \gamma^{\ast} t} 
   \right\} e^{i \omega t} ,
\]
and obtain the linear eigenvalue problem for $v(x)$ and $w(x)$,
\begin{equation} \label{eq:Leigen}
  \begin{array}{l}
   { \displaystyle
      -(\omega+\gamma) v + \frac{d^2 v}{d x^2} + \nu(x) v 
      + \delta(x) \left[ G_1 v + (G_1 - G_0) w \right] = 0 , 
   } \\*[9pt] { \displaystyle
      -(\omega-\gamma) w + \frac{d^2 w}{d x^2} + \nu(x) w 
      + \delta(x) \left[ G_1 w + (G_1 - G_0) v \right] = 0 ,
   } \end{array}
\end{equation}
where $G_1 \equiv G_0 + I_0 G^{\prime}(I_0)$, the intensity $I_0$ is calculated for an unperturbed solution, and the prime stands for the derivative. 

We find that the localized eigenmode solutions of Eq.~(\ref{eq:Leigen}) exist only for the particular eigenvalues satisfying the solvability condition, $Y( \gamma ) = 0$, where $Y( \gamma )$ is the so-called {\em Evans function},
\[
Y( \gamma  ) = 
   \left[ G_1 - \zeta(\omega+\gamma) \right]
   \left[ G_1 - \zeta(\omega-\gamma) \right]
   - \left( G_1 - G_0 \right)^2 . 
\]
In general, the localized eigenmodes fall into one of the following categories:
(i)~{\em internal modes} with real eigenvalues describe periodic oscillations (``breathing'') of the localized state, (ii)~{\em instability modes} correspond to purely imaginary eigenvalues, and (iii)~{\em oscillatory instabilities} can occur when the eigenvalues are complex.
To conduct the further analysis in more details, we need to know the properties of the wave spectrum defined by the type of a linear (homogeneous or periodic) medium.

\section{Impurity-induced localization in~a homogeneous medium}

In order to provide a link with earlier studies, first we apply the general analytical technique to describe the properties of nonlinear impurity modes in a homogeneous medium, i.e. when $\nu(x) \equiv 0$. In this case, we have $\mu = \sqrt{\omega}$, and the Bloch-type solutions correspond to non-interacting counter-propagating waves, with constant values of $a(x)$ and $b(x)$. Therefore, we have $r^{\pm} \equiv 0$, ${\zeta^\pm} = 2 \mu$, and $\tau^\pm = e^{-\mu h}$. Then, the condition $|\tau| \ne 1$ reduces to a simple relation $\omega > 0$, so that $\mu$ is real, which means that {\em wave localization is possible due to the total internal reflection (IR)} only.

To demonstrate the basic properties of the IR mode, we consider a localized impurity possessing a cubic nonlinear response,
\[
  G(I) = \alpha + \beta I.
\]
Under the proper scaling, the absolute value of the nonlinear coefficient~$\beta$ can be normalized to unity, so that $\beta=+1$ corresponds to {\em self-focusing}, and $\beta=-1$ to {\em self-defocusing} nonlinearity. Localization depends also on the sign of the linear coefficient~$\alpha$, which defines the impurity response at small intensities: {\em attractive}, if $\alpha>0$, and {\em repulsive}, otherwise. In this case, the matching condition~(\ref{eq:zeta}) can be readily solved, and we obtain an explicit solution for the profile of the localized waves,
\[
   u(x) =  u_0 e^{-\sqrt{\omega} |x|},  \quad  
   u_0^{2} = \frac{1}{\beta} (2 \sqrt{\omega} - \alpha) ,
\]
and find the corresponding expression for the power,
   $P(\omega) = {u_0^2} / {\sqrt\omega}$.
We can also present the Evans function in a simple form,
\begin{equation} \label{eq:Evans_homog}
  Y(\gamma) = 2 (\eta - 2 \sqrt{\omega}) (\eta - 2 \sqrt{\omega} + \alpha),
\end{equation}
where $\eta = \sqrt{\omega+\gamma} + \sqrt{\omega-\gamma}$, and the square root values are taken on a branch with a positive real part.

\begin{figure}[t]
\setlength{\epsfxsize}{10cm}
\vspace*{-2mm}
\centerline{\mbox{\epsfbox{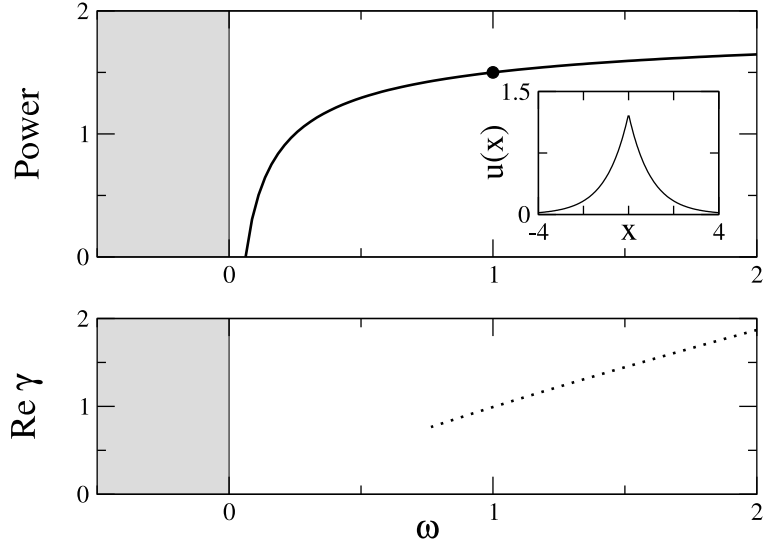}}}
\vspace*{1mm}
\caption{ \label{fig:pwr-homog}
Top: power vs. frequency dependences for the localized waves in a linear homogeneous medium; inset shows the wave profile corresponding to a marked point.
Bottom: real eigenvalue corresponding to the internal mode.
White and gray backgrounds mark the band gap and band, respectively.
The impurity characteristics are $\alpha = 0.5$ and $\beta=1$.}
\end{figure}

Our analysis shows that the localized modes can exist in the presence of self-defocusing nonlinearity ($\beta<0$) only for $\alpha>0$. Then, the family of localized solutions bifurcates from the linear limit at the {\em cut-off frequency} $\omega_b=\alpha^2/4$, with $P(\omega_b)=0$, and exists for $0<\omega<\omega_b$. It is straightforward to check that the equation $Y(\gamma)=0$, for the Evans function given by Eq.~(\ref{eq:Evans_homog}), has solutions only in the trivial case $\gamma=0$ (corresponding to $\eta=2\sqrt{\omega}$), since a formal solution $\eta = (2\sqrt{\omega}-\alpha)=2(\sqrt{\omega}-\sqrt{\omega_b})$ is negative, and thus it can not correspond to any $\gamma$. Therefore, the nonlinear states supported by a nonlinearly repulsive impurity are stable.

In the case of self-focusing nonlinearity ($\beta>0$), the modes appear above the frequency cut-off, $\omega>\omega_b$, if $\alpha>0$, and they exist for $\omega>0$, provided $\alpha<0$. Non-trivial linear eigenvalues exist if $\sqrt{\omega} > \alpha / (2-\sqrt{2})$, and they are given by the following expression,
$\gamma = \pm 2 (\sqrt{\omega}-\alpha/2) 
              [ \alpha (\sqrt{\omega} - \alpha/4)]^{1/2}$. It then follows that $\gamma$ is imaginary, and the modes are {\em unstable}, if $\alpha < 0$, and {\em stable} if $\alpha>0$, when $\gamma$ is real. We would like to note that for the self-focusing nonlinearity the mode stability can also be determined using the {\em Vakhitov-Kolokolov criterion}~\cite{VK}, which states that the modes are stable, if $d P / d \omega > 0$, and unstable, otherwise. Indeed, the slope of the dependence $P(\omega)$ is always positive if $\alpha > 0$, as illustrated in Fig.~\ref{fig:pwr-homog}.

\section{Impurity-induced localization in~a periodic superlattice}

\begin{figure}[b]
\setlength{\epsfxsize}{10cm}
\vspace*{-2mm}
\centerline{\mbox{\epsfbox{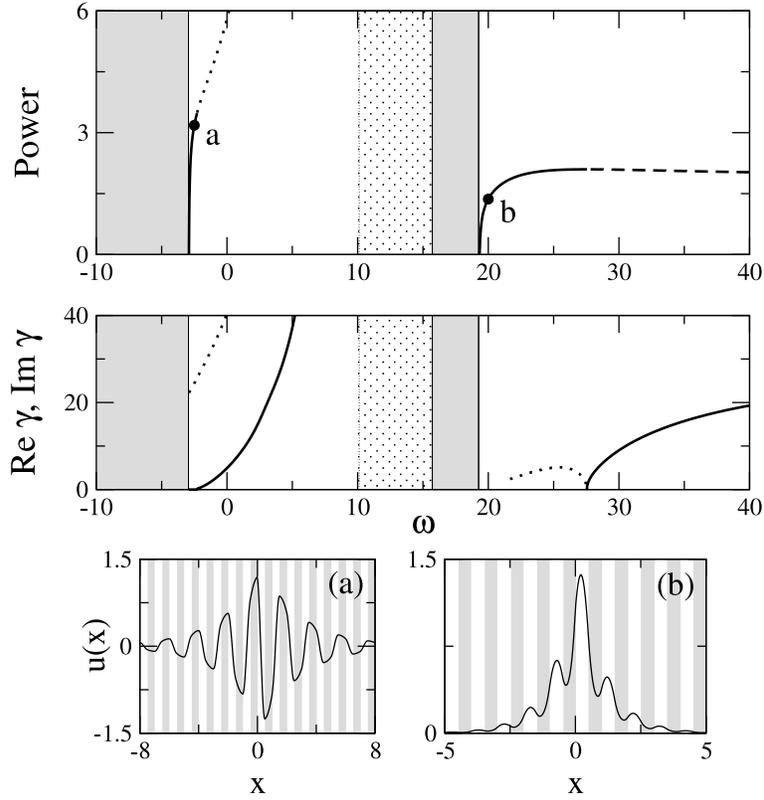}}}
\vspace*{1mm}
\caption{ \label{fig:pwr-sf-alp} \label{fig:bandgap}
Top: power vs. frequency dependences for the localized waves: solid~--- stable, dashed~--- unstable, and dotted~--- oscillatory unstable.
Middle: real (dotted) and imaginary (solid) parts of the eigenvalues associated with the wave instability.
Shading marks ``waveguiding'' (white) and ``anti-waveguiding'' (dotted) localization regimes inside the band gaps.
Bottom:~the localized states corresponding to the marked points (a,b) in the top plot; shading marks the areas with smaller $\nu$.
The lattice parameters are $h=1$, $\nu(x)=0$ for $n-1/2 < x/h < n$, and $\nu(x)=30$ for $n < x/h < n+1/2$, where $n$ is integer.
The impurity characteristics are $\alpha = 0.5$ and $\beta=1$.}
\end{figure}

We now consider the properties of the localized waves supported by an impurity with a self-focusing nonlinearity ($\beta=+1$), and attractive linear response ($\alpha>0$), in a two-component superlattice. Due to a periodic modulation of the effective potential, $\nu(x)$, such a lattice possesses {\em several band gaps}, as demonstrated in Fig.~\ref{fig:bandgap} (top). 

The first (semi-infinite) band gap exists due to {\em the internal reflection} (IR) and, therefore, the properties of the corresponding localized waves should be similar to those existing in a homogeneous medium. Indeed, we find that the localized states in the IR band gap resemble conventional impurity modes (or solitary waves) modulated by a periodic structure [see Fig.~\ref{fig:bandgap}(b)], and these states exist above the cut-off frequency $\omega_b$ defined by the equality $\zeta(\omega_b^{\rm (IR)})=\alpha$. Since such an IR wave is a fundamental eigenstate of the self-induced waveguide, it can be demonstrated that the conditions of the Vakhitov-Kolokolov (VK) stability theorem~\cite{VK} are satisfied, and the IR states are {\em unstable} if and only if $d P / d \omega < 0$. The critical point $d P / d \omega = 0$ corresponds to a ``collision'' between two internal modes of the localized wave at the origin, as illustrated in Fig.~\ref{fig:pwr-sf-alp} (middle). It is interesting to note that the high-frequency localized waves can become unstable in a periodic superlattice, although they are linearly stable when a nonlinear impurity with similar characteristics is embedded in a homogeneous medium (cf. Figs.~\ref{fig:pwr-homog} and~\ref{fig:pwr-sf-alp}).

\begin{figure}[t]
\setlength{\epsfxsize}{10cm}
\centerline{\mbox{\epsfbox{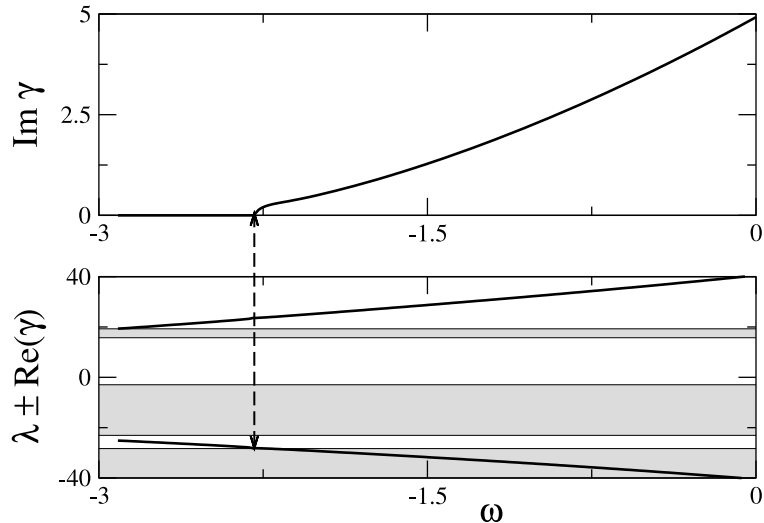}}}
\caption{ \label{fig:imode-sf-alp}
An example of a resonance that occurs between an internal mode of the localized wave and a bang-gap edge, leading to an oscillatory instability.}
\end{figure}

Additional band gaps appear at lower frequencies due to the resonant {\em Bragg-type reflection} (BR) from a periodic structure. The corresponding localized waves exist in the waveguiding regime (white regions in Fig.~\ref{fig:pwr-sf-alp}, top), above the cut-off frequencies $\omega_b^{\rm (BR)}$, where again $\zeta(\omega_b^{\rm (BR)})=\alpha$. Such localized solutions are somewhat similar to the gap solitons composed of mutually coupled backward and forward propagating waves [see Fig.~\ref{fig:bandgap}(a)]. For the BR states, the VK criterion provides only a necessary condition for stability. Indeed, we notice that in the linear limit there always exists an internal mode corresponding to a resonant coupling between the BR and IR band-gaps, since $Y \left( \omega_b^{\rm (IR)}-\omega_b^{\rm (BR)} \right) \equiv 0$. We perform extensive numerical calculations and find that this mode leads to an oscillatory instability of BR waves when the value $(\omega-{\rm Re}\;\gamma)$ moves outside the band gap; it happens when the intensity exceeds a threshold value (see Figs.~\ref{fig:pwr-sf-alp} and~\ref{fig:imode-sf-alp}).

\section{Concluding remarks}

We have analyzed the existence and stability of nonlinear localized waves in one-dimensional periodic structures (or superlattices) considering the simplest nonlinear generalization of the Kronig-Penney model with a nonlinear impurity. Taking into account many common features of nonlinear guided waves and impurity modes in stratified and disordered media, on one side, and the self-trapped states and solitary waves in homogeneous nonlinear media, on the other side, we expect that many of the results described above will be found in more realistic physical models of nonlinear periodic media. Such cases include the electron self-trapping and locking states in cuprates and semiconductor superlattices~\cite{a1}; nonlinear guided waves in optical superlattices~\cite{a2}; impurity modes in photonic band-gap materials~\cite{a3}, magneto-optical periodic structures~\cite{a4}, and photonic crystal fibers~\cite{pcf}; coherent matter waves in optical lattices~\cite{a5}, etc. In particular, stability of nonlinear localized impurity modes is a crucial issue for creating tunable band-gap materials where gaps could be controlled by changing the input light intensity.

We are indebted to O.~Bang for useful collaboration at the initial stage of this project.
The work has been partially supported by the Performance and Planning Fund of the Institute of Advanced Studies.

\end{sloppy}

\begin{chapthebibliography}{1}
\bibitem{ref}
Using the optics terminology, we notice that such two types of localized states are the guided waves corresponding to the conventional waveguiding, due to the total internal reflection, and to the band-gap states, due to the Bragg-type reflection.
\end{chapthebibliography}
\end{document}